\documentclass{pos}
\newcommand{\beao}{\begin{eqnarray*}}
\newcommand{\eeao}{\end{eqnarray*}}
\newcommand{\bea}{\begin{eqnarray}}
\newcommand{\eea}{\end{eqnarray}}
\newcommand{\be}{\begin{equation}}

\newcommand{\ee}{\end{equation}}

\newcommand{\hk}{\hspace{0.1cm}}

\newcommand{\rk}{\right)}
\newcommand{\lk}{\left(}

\newcommand{\sli}{\sum\limits}

\renewcommand{\vec}[1]{\mbox{\boldmath$#1$\unboldmath}}

\newcommand{\vB}{{\vec{B}}}

\title{Recent results from the Hamiltonian approach\\to Yang--Mills theory in Coulomb gauge}

\ShortTitle{Coulomb
gauge Yang--Mills theory }

\author{\speaker{H. Reinhardt}\thanks{supported by DFG-Re 856/6-2,3}, G . Burgio, D. Campagnari, D. Epple, M. Leder, 
M. Pak, M. Quandt and W.
Schleifenbaum\\
        University of T\"ubingen \\      
            Institute of Theoretical Physics\\
	    Auf der Morgenstelle 14\\
	    D-72076 T\"ubingen\\
        Email: \email{hugo.reinhardt@uni-tuebingen.de}}

\abstract{Within the Hamiltonian approach to Yang-Mills theory
 in Coulomb gauge the ghost and gluon
propagators are determined from a variational solution of the
Yang--Mills Schr\"odinger equation  showing both
gluon and heavy quark confinement. The continuum results  are in good agreement
with lattice data. The ghost form factor is identified as the
dielectric function of the Yang--Mills vacuum and
the Gribov--Zwanziger confinement scenario is shown to imply the dual Meissner
effect. The topological susceptibility is calculated.}

\FullConference{8th Conference Quark Confinement and the Hadron Spectrum \\
		 September 1-6 2008\\
		 Mainz, Germany}

\begin{document}

\begin{figure}
  \centerline{
  \includegraphics{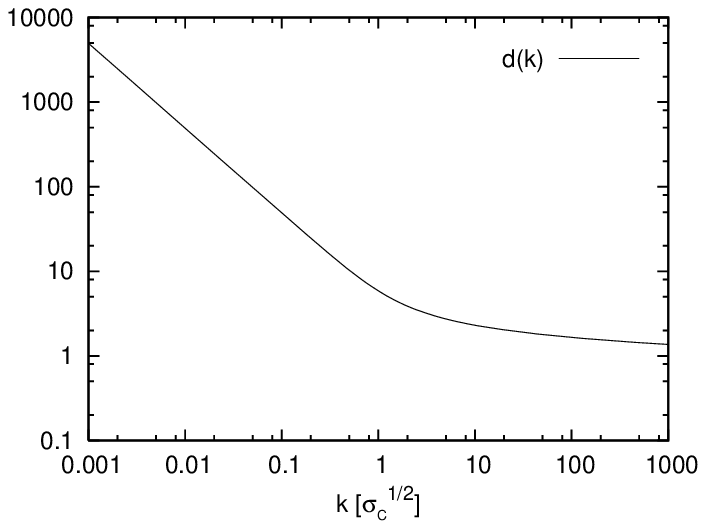}
  \includegraphics{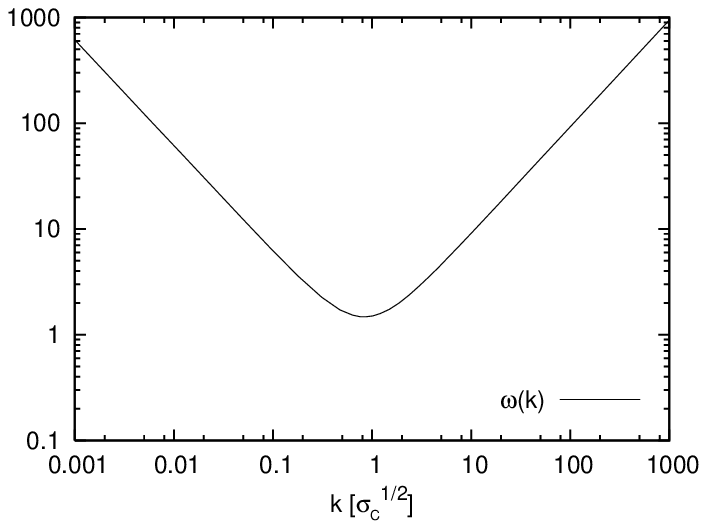}}
  \caption{Ghost form factor $d(k)$ (left) and gluon energy $\omega(k)$ from the variational solutions presented in \cite{EppReiSch07}.}
  \label{d+w}
\end{figure}
We  solve the Schr\"odinger equation $H \psi = E \psi$ of Yang-Mills theory in
Coulomb gauge 
by the variational principle $\langle \psi | H | \psi \rangle \to min$
with the following ansatz for the vacuum wave functional $\psi (A^\perp)$
\cite{FeuRei04}. 
\be
\label{G15}
\psi (A^\perp) = \frac{1}{\sqrt{J (A^\perp)}} \exp \lk - \frac{1}{2} \int d^3 x
d^3 y A^{\perp a}_i (x) \omega (x, y) A^{\perp a}_i (y) \rk \hk ,
\ee
where the kernel $\omega (x, y)$ is determined from the variational
principle \cite{EppReiSch07}, \cite{SzcSwa02}, \cite{FeuRei04} 
and $J (A^\perp) = Det (- \hat{D}
\partial)$ is the Faddeev-Popov determinant. In practice the so resulting 
equation for $\omega (x, y)$ is
converted into a set of Dyson-Schwinger equations for the gluon propagator
\be
\label{G16}
\langle A^{\perp a}_i (x) A^{\perp b}_j (y) \rangle = \delta^{a b} t_{i j}(x) \frac{1}{2}
\omega^{- 1} (x, y) \; ,
\ee
with $t_{ij}(x)=\delta_{ij}-\frac{\partial_i\partial_j}{\partial^2}$
being the transverse projector,
and the ghost propagator
\be
\label{G17}
G (x, y) = \left\langle \lk - \vec{\hat{D}}\cdot \vec{\partial} \rk^{-1} \right\rangle
=  \langle x |d (-
\Delta) (- \Delta)^{- 1} | y \rangle \hk .
\ee
Here we have introduced the ghost form factor $d (- \Delta)$, which describes
the deviation of the QCD ghost propagator from the QED case, where $d (- \Delta)
\equiv 1$. The resulting Dyson-Schwinger equations need renormalisation, which
is well under control \cite{RX}. Fig.\ \ref{d+w} shows the solution of the Dyson-Schwinger equation
for the gluon energy $\omega (k)$ and the ghost form factor $d (k)$,
as shown in Ref.\ \cite{EppReiSch07}. 
An analytic
infrared and ultraviolet analysis of the Dyson-Schwinger equation shows the
following asymptotic behaviour \cite{FeuRei04,SchLedRei06}
\bea
\label{G18}
\mathrm{IR}\, (k \to 0) & : & \omega (k) \sim \frac{1}{k} \hspace{1cm} d (k) \sim
\frac{1}{k} \nonumber\\
\mathrm{UV}\, (k \to \infty) & : & \omega (k) \sim {k} \hspace{1cm} d (k) \sim k^0
\hk .
\eea
At large momenta the gluon behaves like a photon, which is in agreement with
asymptotic freedom, while at small momenta the gluon energy diverges, which
implies the absence of gluon states in the physical spectrum. This is nothing
but a manifestation of gluon confinement. The infrared divergence of the ghost
form factor is a consequence of the horizon condition 
\be
\label{G19}
d^{- 1} (k = 0) = 0  \hk ,
\ee
which has been used as input in the renormalisation of the ghost Dyson-Schwinger
equation. This is a necessary condition for the Gribov-Zwanziger confinement
scenario. In fact, one can show that there is a sum rule relating the infrared
exponents of the ghost and the gluon propagator and an infrared divergent gluon
energy requires also an infrared divergent ghost form factor, i.e.\ the horizon
condition (\ref{G19}), see Ref.\ \cite{SchLedRei06}. A similar behaviour of the
propagators is also obtained from functional renormalization group 
flow equations \cite{RX1}. 
\begin{figure}
  \centerline{
  \includegraphics{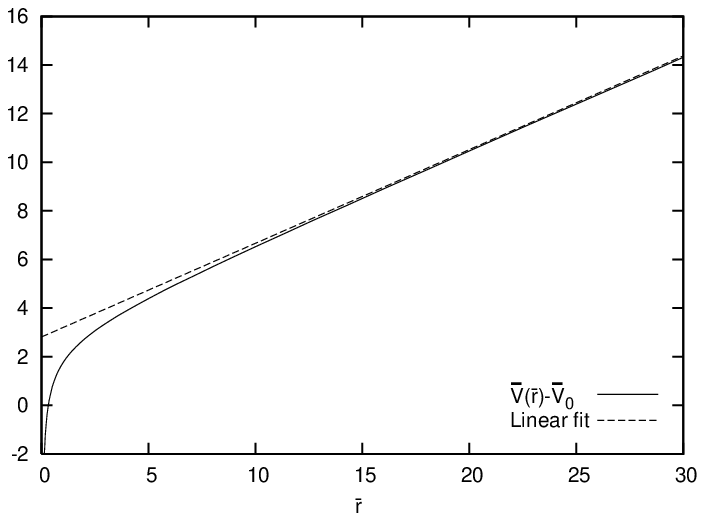}
  \includegraphics{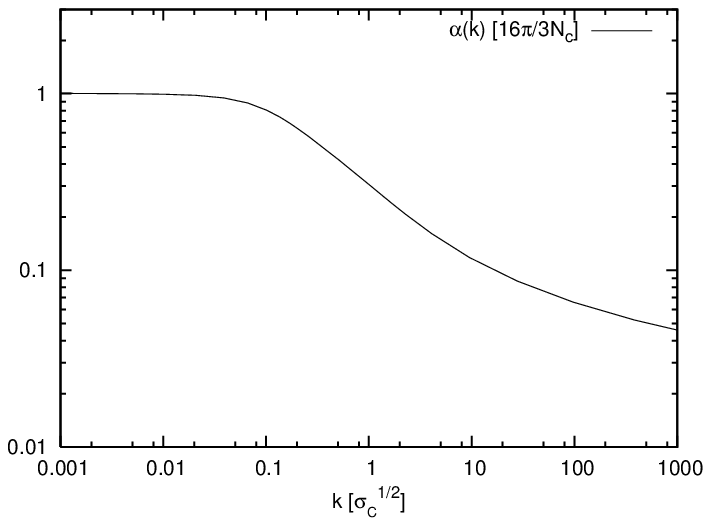}}
  \caption{\label{Vc+alpha}\textit{Left:} Heavy quark potential given by eq.
           {(\protect\ref{G20})}. \textit{Right:} Running coupling constant.}
\end{figure}
Fig.\ \ref{Vc+alpha} shows the non-Abelian Coulomb potential
\be
\label{G20}
V (| x - y|) = g^2 \left\langle \langle x |
 ( - \vec{\hat{D}}\cdot \vec{\partial} )^{-1}
  (- \vec{\partial}^2) (- \vec{\hat{D}}\cdot \vec{\partial})^{- 1} | y
  \rangle \right\rangle \stackrel{|x - y| \to \infty}{\longrightarrow}
   \sigma_C\, |x-y|
\hk ,
\ee
which for large distances indeed increases linearly \cite{EppReiSch07} as the infrared analysis
reveals. The Coulomb string tension $\sigma_C$ sets the scale of our
approach. Also shown in Fig.\ \ref{Vc+alpha} is the running coupling constant which
is infrared finite, for details see Ref.\ \cite{SchLedRei06}.

In $D = 3 + 1$ dimensions, previous lattice calculations performed in Coulomb gauge in
Ref.\ \cite{LanMoy04, Qua+07} showed an anomalous UV behaviour of the gluon
propagator --- $\mathrm{IR}: \omega (k) \sim k^0 \, ,\; \mathrm{UV}: \omega (k) \sim
k^{3/2}$ --- which is in strong conflict with the continuum result. However, one
should mention that these lattice calculations assumed multiplicative
renormalisability of the 4-dimensional gluon propagator, which give rise to
scaling violations in the static propagator. Furthermore, these calculations did
not fix the gauge completely, i.e.\ the residual time-dependent gauge invariance
left after Coulomb gauge fixing was left unfixed. 

\begin{figure}
  \centerline{
  \includegraphics[scale=1.0]{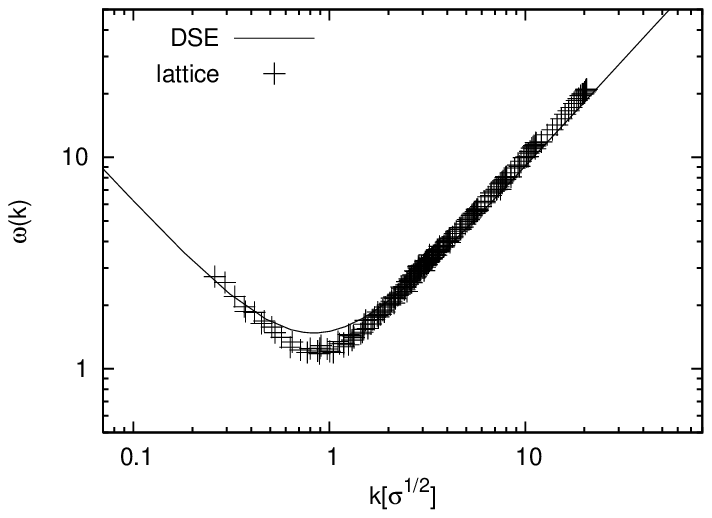}
  \includegraphics[scale=1.0]{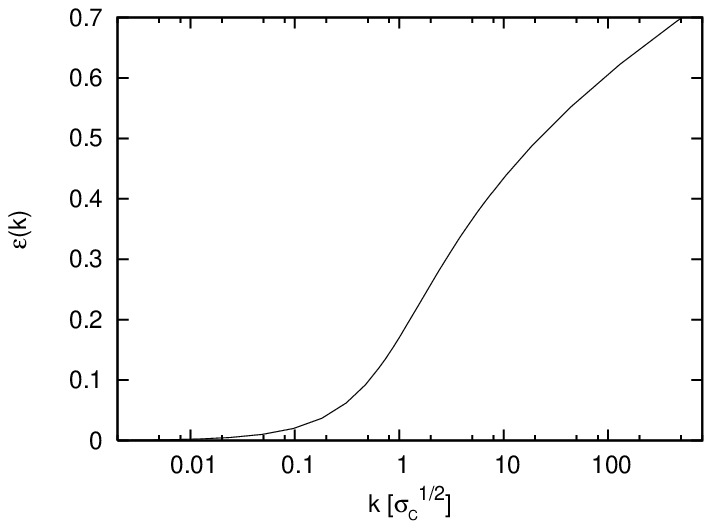}}
  \caption{{\it Left:} Lattice data for $\omega(k)$, compared to the
    solution of   the Dyson-Schwinger equations.  {\it Right:} Dielectric function $\epsilon(k)$.}
  \label{latt+eps}
\end{figure}

Recently, we have done improved lattice calculations with a complete gauge
fixing \cite{Bur+08}. In these studies, the energy dependence of the 4-dimensional gluon propagator
could be explicitly extracted and it was found that the static gluon propagator
is multiplicatively renormalisable and shows a perfect scaling. 
Fig.\ \ref{latt+eps} (left panel) shows the
results for the gluon propagator of these calculations together with the
continuum results. It is assumed here that the Coulomb string tension
$\sigma_C$ is identical to the string tension $\sigma$ from the Wilson
loop. There is a good agreement, particularly in the  infrared and ultraviolet
 the lattice and continuum results match perfectly. It is also
remarkable that the lattice result can be very well fitted by Gribov's original
formula for the gluon energy
\be
\label{21}
\omega (k) = \sqrt{k^2 + \frac{M^4}{k^2}}
\ee
with $M = 0.88(1)\,\mathrm{GeV}$.

In ref. \cite{Rei08} it was shown that 
the inverse of the ghost form factor $d(k)$ can be identified as the dielectric
function of the Yang-Mills vacuum
\be
\label{G24}
\epsilon (k) = d^{- 1} (k) \hk .
\ee
Fig.\ \ref{latt+eps} (right panel) shows the so defined dielectric function. It satisfies $0 < \epsilon (k)
< 1$, which is a manifestation of anti-screening while in QED we have $\epsilon
(k) > 1$, which corresponds to ordinary Debye screening. Furthermore, at zero
momentum the dielectric function vanishes, showing that in the infrared the
Yang-Mills vacuum behaves like a perfect colour dia-electric medium. The vanishing of
the dielectric function in the infrared is not an artifact of our solutions of
the Dyson-Schwinger equations but is guaranteed by the horizon condition, which
is a necessary condition for the Gribov-Zwanziger confinement scenario. A
perfect colour dia-electric medium $\epsilon = 0$ is nothing but a dual
superconductor. (Here, ``dual'' refers to an interchange of electric and magnetic
fields and charges.) Recall in an ordinary superconductor the magnetic
permeability vanishes $\mu = 0$. This shows that the Gribov-Zwanziger
confinement scenario implies the dual Meissner effect \cite{Rei08}.

\begin{figure}
  \centering
  \includegraphics[scale=0.6]{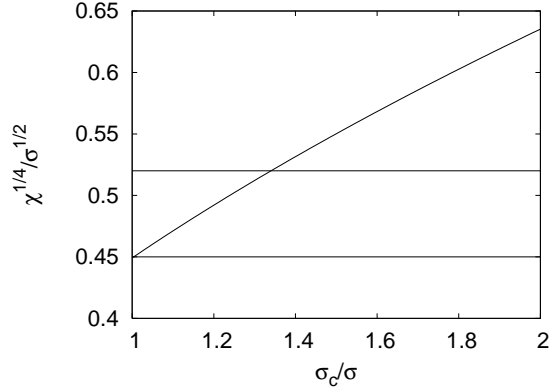}
  \caption{Topological susceptibility $\chi$ as a function of the ratio
 $\sigma_c /\sigma$.}
  \label{chi}
\end{figure}

 In the Hamiltonian approach one finds the
following expression for the topological susceptibility \cite{CamRei08}
(V: spatial volume)
\be
\label{G28}
V \chi = \lk \frac{g^2}{8 \pi^2} \rk^2 \left[ \langle 0 | \int \vB^2 (x) | 0
\rangle - 2 \sli_n \frac{| \langle n | \int \vB \cdot\vec{\Pi} | 0 \rangle |^2}{E_n}
\right] \hk .
\ee
Here $| n \rangle$ denotes the exact excited states of the Yang-Mills
Hamiltonian with energies $E_n$. These eigenstates are of course not known. We
work out the matrix elements in eq.\ (\ref{G28}) to two-loop order. In this order
only two- and three-quasi gluon states 
contribute where the quasi gluons are defined as excitations of the vacuum
$\langle 0 | A \rangle = \Psi (A)$ (\ref{G15}) with energy $\omega (k)$. 
The resulting expression for the topological
susceptibility is ultraviolet divergent and needs renormalisation. For this aim
we exploit the fact that $\chi$ vanishes to all order perturbation theory and
renormalise the expression (\ref{G28}) for $\chi$
by subtracting each propagator by its perturbative
expression. This renders $\chi$ (\ref{G28}) finite. Furthermore, since the
momentum integrals in this expression are dominated by the infrared part we
replace the coupling constant, which, in principle, should be the running one,
 by
its infrared value. The results obtained in this way for the topological
susceptibility are shown in Fig.\ \ref{chi} (right panel) as a function of the ratio $\sigma_C / \sigma$. Choosing $\sigma_C = 1.5 \sigma$ 
which is the value favoured
by the lattice calculation \cite{LanMoy04} we find for $SU (2)$ with
$\sqrt{\sigma} = 440 \hk MeV$
\be
\label{G30}
\chi = (240 \hk  MeV)^4 \hk .
\ee
This value is somewhat larger than the lattice prediction $\chi = (200 - 230 \hk 
MeV)^4$.

The results obtained so far in the present
approach are quite encouraging for further investigations. A natural next step
would be the inclusion of dynamical quarks.


\end{document}